\documentclass{article}
\usepackage{graphicx}
\graphicspath{%
    {converted_graphics/}
    {/}
    {C:/Dropbox/1-kgl-top/Papers/2016/SM5-MeasAt-5C/}
}
\begin{document}

\begin{center}
{\LARGE Toward a Comprehensive Model of Snow Crystal Growth:}\vskip6pt

{\LARGE \ 5. Measurements of Changes in Attachment Kinetics}\vskip6pt

{\LARGE from Background Gas Interactions at -5 C}\vskip16pt

{\Large Kenneth Libbrecht}\vskip4pt

{\large Department of Physics, California Institute of Technology}\vskip1pt

{\large Pasadena, California 91125}\vskip4pt

\vskip18pt

\hrule\vskip1pt \hrule\vskip14pt
\end{center}

\textbf{Abstract.} We present measurements of the diffusion-limited growth
of ice crystals from water vapor at a temperature of -5 C, in air at a
pressure of $p_{air}=1$ bar. Starting with thin, c-axis ice needle crystals,
the subsequent growth morphologies ranged from solid prismatic columns to
hollow columns to complex \textquotedblleft fishbone\textquotedblright\
dendritic structures as the supersaturation was increased. We modeled the
simpler morphologies using analytical techniques together with a
cellular-automata method that yields faceted crystalline structures in
diffusion-limited growth. We found that the molecular attachment coefficient 
$\alpha _{prism}$ on faceted prism surfaces in air at -5 C is substantially
lower than that measured at low background air pressure. Our data show that
increasing $p_{air}$ from 0.01 to 1 bar reduces $\alpha _{prism}$ by nearly
two orders of magnitude at this temperature. In contrast, we find that $%
\alpha _{basil}$ is essentially unaffected by air pressure over this range.
These and other measurements indicate that ice surfaces near the melting
point undergo a series of complex structural and dynamical changes with
temperature that remain largely unexplained at even a qualitative level.

\section{Introduction}

Our overarching goal in this series of investigations is to develop a
comprehensive model of ice crystal growth from water vapor that can
reproduce quantitative growth rates as well as growth morphologies over a
broad range of circumstances. Although ice crystal formation has been
studied extensively for many decades, our understanding of the fundamental
physical processes governing growth behaviors at different temperatures and
supersaturations remains remarkably poor \cite{nakaya54, mason63, lamb72,
kurodalac82, lacmann83, nelsonknight98, libbrechtreview05}. For example, the
complex dependence of ice growth morphology on temperature, exhibiting
several transitions between plate-like and columnar structures \cite%
{morph04, hallet09}, remains essentially unexplained even at a qualitative
level, although it was first reported over 75 years ago \cite{nakaya54}.

To address this problem, we have undertaken an experimental program designed
to create small ice crystals with simple morphologies and measure their
subsequent growth under carefully controlled conditions, to an extent and
accuracy surpassing previous efforts \cite{knight12, hallet12, maruyama05,
nelson01, nelsonknight98, kuroda84}. We model the experimental data using a
recently developed cellular-automata numerical method that can generate
physically realistic faceted structures in diffusion-limited growth \cite%
{reiter05, gg08, gg09, kglca13, kglSDmodel, kelly13}. The comparison between
measured and modeled ice crystals then provides valuable information about
the attachment kinetics governing ice growth from water vapor. From this
information we hope to develop a detailed physical picture of the molecular
structure and dynamics of the ice surface during solidification.

\section{Ice Growth Measurements in a Dual Diffusion Chamber}

The ice growth measurements described in this paper were obtained used the
dual diffusion chamber described in \cite{kgldual14}. The first of the two
diffusion chambers was operated with a high water-vapor supersaturation in
air, and in this chamber we grew electrically enhanced ice needles with tip
radii \symbol{126}100 nm and overall lengths up to 4 mm, with the needle
axis oriented along the c-axis of the ice crystal. The needle crystals were
then transported to the second diffusion chamber, also operating in air at $%
p_{air}=1$ bar, where the temperature and supersaturation were independently
controlled, and the subsequent growth was recorded using optical microscopy.
A well-defined linear temperature gradient in the second chamber ensured
that convection currents were suppressed and that the supersaturation could
be accurately modeled. In \cite{kglcyl} we describe the supersaturation
modeling in more detail, along with a calibration of the supersaturation at
the location of the test crystals.

\begin{figure}[htb] 
  \centering
   \includegraphics[width=5.5in,keepaspectratio]{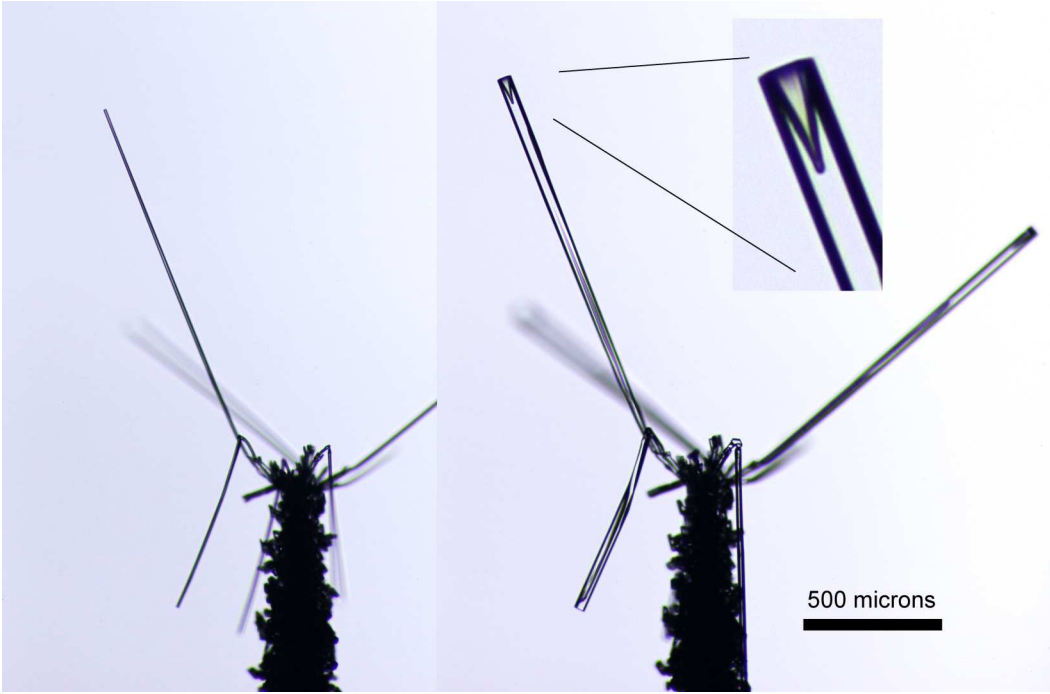}
  \caption{This composite photograph shows
an example of a hollow columnar crystal grown at -5 C. The image in the left
panel was taken soon after several thin, c-axis \textquotedblleft
electric\textquotedblright\ ice needles had grown out from the wire
substrate covered in frost crystals. Focusing on a single needle, the right
panel shows its structure after an additional nine minutes of growth at a
supersaturation of $\protect\sigma _{center}\approx 3.7$ percent (for this
particular example). The magnified inset image shows the columnar hollowing
that developed as the crystal grew. The diameter and length of the ice
column as a function of time were extracted from a set of similar images.}
  \label{example}
\end{figure}

Immediately after an ice needle assembly was moved to the second diffusion
chamber, the wire base holding the needles was rotated so a particular test
needle was in focus in the microscope, with the needle entirely in the focal
plane, as shown in Figure \ref{example}. During this transport and focusing
step, a thin, frost-covered, horizontal shutter plate was positioned just
above the ice needles, reducing the supersaturation below the plate to near
zero. Once the test needle was satisfactorily positioned (typically taking
10-30 seconds), the shutter was removed and growth measurements commenced.
The supersaturation near the test crystal relaxed back to steady state in a
time of order $\tau \approx L^{2}/D\approx 5$ seconds, where $L\approx 1$ cm
is the shutter size and $D\approx 2\times 10^{-5}$ m$^{2}/$sec is the
diffusion constant for water vapor in air.

For all the measurements described below, the air temperature was maintained
at $T_{center}=-5\pm 0.1$ C, as determined by a small calibrated thermistor
that was frequently placed at the center of the diffusion chamber, at the
same location as the tips of the ice needles during their growth. The
supersaturation was adjusted by changing the linear temperature gradient
inside the diffusion chamber. In particular, the top and bottom temperatures
were maintained at $T_{center}\pm \Delta T$, and the supersaturationr was
proportional to $\Delta T^{2}$, as described in \cite{kgldual14, kglcyl}.
The supersaturation calibration presented in \cite{kglcyl} gives $\sigma
_{center}\approx 0.00148\left( \Delta T\right) ^{2}$ when $T_{center}=-5$ C,
where $\sigma _{center}$ is the supersaturation far from the growing ice
crystals. The supersaturation $\sigma _{surface}$ at the ice surface must be
determined by diffusion modeling.

In a typical growth run at $-5$ C (measuring a single needle crystal), still
photos were taken periodically to record the growth after the shutter was
removed. For supersaturations $\sigma _{center}<2$ percent surrounding the
growing crystals, ice needles grew slowly into solid, prismatic columnar
structures. The morphology changed with increasing $\sigma _{center},$
yielding predominantly hollow columns at $\sigma _{center}\approx 4$
percent, as shown in Figure \ref{example}. At still higher $\sigma _{center},
$ the corners of a hollow column separated to yield needle-like crystals,
and a sextet of \textquotedblleft fishbone\textquotedblright\ crystals \cite%
{fishbones09, kgldual14} appeared when $\sigma _{center}>15$ percent. 

The work presented here is limited to $\sigma _{center}<4$ percent, so the
growth morphologies all exhibited solid or hollow columnar morphologies. The
diffusion-limited growth of these structures could be quantitatively modeled
using a 2D cylindrically symmetric cellular automata code, as described in 
\cite{kglca13}, thus avoiding the need for a full 3D code.

The columnar radius at the end of an ice needle as a function of time, $R(t),
$ was extracted directly from the image data. The optical microscope used to
photograph the crystals had a resolving power of 2.5 $\mu $m, and the image
pixels measured 0.85 $\mu $m. Our diameter resolution was found to be about $%
\pm 2$ $\mu $m, giving measurements of $R(t)$ that were accurate to about $%
\pm 1$ $\mu $m. We did not distinguish the different \textquotedblleft
radii\textquotedblright\ of a projected hexagonal structure in our image
data, and this limited the absolute accuracy of our measurements of $R(t)$
to $\pm 5$ percent. Moreover, the faceted needles often did not remain
perfectly hexagonal in cross section as they grew, adding additional
systematic errors arising from our cylindrically symmetric modeling of the $%
R(t)$ data.

The height of a needle, $H(t),$ was measured with respect to a
\textquotedblleft base reference\textquotedblright\ that consisted of one or
more reference points in the frost cluster covering the wire substrate at
the base of the needle (for example, see Figure \ref{example}). The quality
and stability of the base references varied from run to run, and the
accuracy of the $H(t)$ measurements was limited to about $\pm 2$ $\mu $m by
growth of the ice crystals in the base reference with time.

With our dual chamber apparatus, we observed the growth of many crystals
over a broad range of $T_{center}$ and $\Delta T,$ including many at $%
T_{center}=-5$ C. Overall, needle crystals grown under similar conditions
yielded similar morphologies and growth data. However, variations in the
exact value of $\Delta T,$ the length and angle of the initial needle, the
needle morphology, the amount of frost on the wire substrate, and the
location of neighboring crystals, resulted in some run-to-run variability in 
$\sigma _{center}$ and other growth parameters. For purposes of clarity,
therefore, we restrict the quantitative analysis described below to just a
few individual crystals of exceptional quality. Data from additional
crystals confirmed our principal conclusions, but are not presented here.

\begin{figure}[htb] 
  \centering
  \includegraphics[width=3.6in,keepaspectratio]{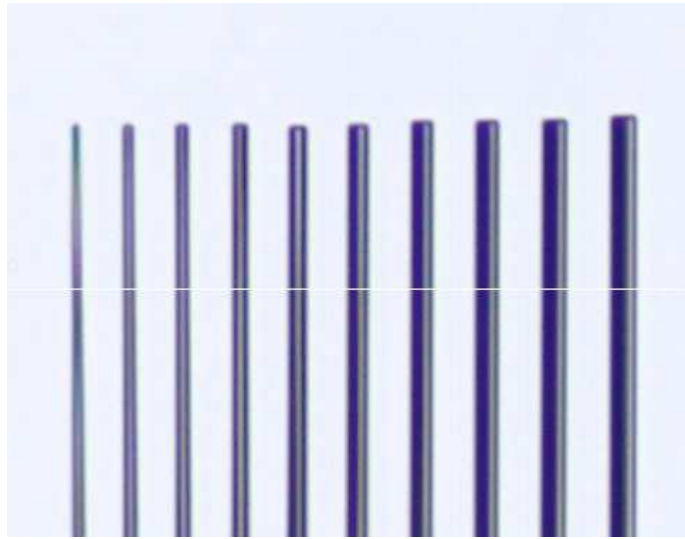}
  \caption{This composite image shows the
growth of a simple columnar ice crystal in air at a supersaturation of $%
\protect\sigma _{center}=0.92$ percent. The columnar radius and height as a
function of time were extracted from these calibrated optical images,
yielding the measurements in Figure \protect\ref{lowsigdata}. (These images
correspond to every other data point in the graph.) A horizontal white line
is drawn 100 $\protect\mu $m below the tips of the first several columns.}
  \label{lowsigimages}
\end{figure}

\subsection{Solid Columnar Growth at Low Supersaturation}

\begin{figure}[htb] 
  \centering
  \includegraphics[width=3.92in,height=3.21in,keepaspectratio]{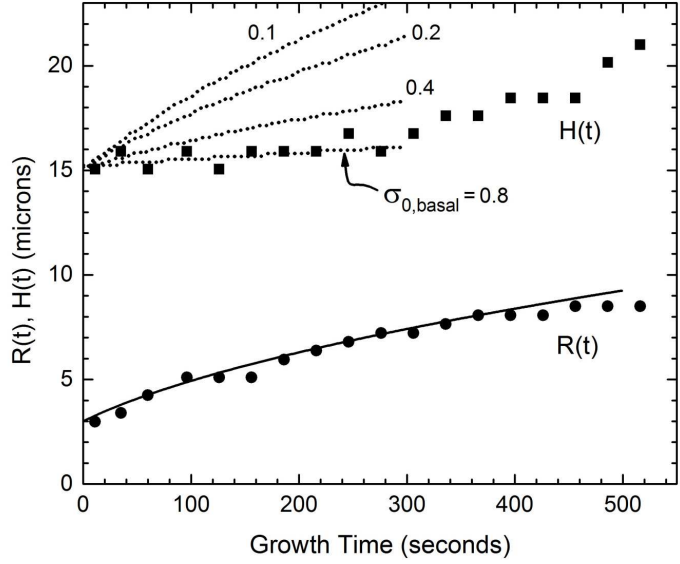}
  \caption{The data points in this graph
show measurements of $R(t)$ and $H(t)$ extracted from image data including
the images shown in Figure \protect\ref{lowsigimages}. A constant length was
subtracted from the $H(t)$ measurements to obtain the (arbitrary) starting
point $H(t=0)\approx 15$ $\protect\mu $m. The various lines are from
diffusion modeling described in the text.}
  \label{lowsigdata}
\end{figure}

Figure \ref{lowsigimages} shows raw imaging data of a solid columnar crystal
as it grew in air with $\Delta T=2.5$ C, for which our supersaturation
calibration yields $\sigma _{center}\approx 0.92$ percent \cite{kglcyl}.
Figure \ref{lowsigdata} shows the measured $R(t)$ and $H(t)$ data obtained
from the full set of images (of which Figure \ref{lowsigimages} shows a
subset), along with model calculations described below.  Close inspection of
the images revealed that the initial needle exhibited a \textquotedblleft
positive\textquotedblright\ taper, with the tip of the needle having the
smallest radius. We measured that the needle sides were tilted with respect
to the c-axis by $dR/dz\approx 0.007$ at $t=0,$ becoming essentially fully
faceted (with the $dR/dz$ measurements extrapolating to zero) by about $%
t\approx 300$ seconds.

To model $R(t),$ we approximate the crystal as an infinitely long cylinder,
which has the growth velocity \cite{kglcyl}%
\begin{equation}
\frac{dR}{dt}=\frac{\alpha _{diffcyl}\alpha _{cyl}}{\alpha _{diffcyl}+\alpha
_{cyl}}v_{kin}\sigma _{center}  \label{diffcyl2}
\end{equation}%
where $\alpha _{cyl}$ gives the attachment coefficient on the cylindrical
surface and 
\begin{equation}
\alpha _{diffcyl}=\frac{1}{B}\frac{X_{0}}{R_{in}},  \label{alphadiff}
\end{equation}%
with $B=\log (R_{far}/R_{in})$ and $X_{0}=c_{sat}D/c_{ice}v_{kin}\approx
0.142$ $\mu $m \cite{kglcyl}. Using $R_{far}=2$ cm and $R_{in}=R(t)\approx 6$
$\mu $m yields $\alpha _{diffcyl}\approx 0.003.$

The needle taper indicates a vicinal surface (inclined slightly from a
faceted prism surface) that includes many molecular steps on the sides of
the needle, so we expect a rather high attachment coefficient $\alpha
_{vicinal}$. Assuming $\alpha _{cyl}=\alpha _{vicinal}\gg \alpha _{diffcyl}$%
, we can write the cylindrical growth rate more simply as  
\begin{equation}
\frac{dR}{dt}=\alpha _{diffcyl}v_{kin}\sigma _{center}  \label{diffcyleqn}
\end{equation}%
which integrates to%
\begin{equation}
R(t)=\left[ \frac{2X_{0}v_{kin}\sigma _{center}}{B}(t-t_{0})+R_{0}^{2}\right]
^{1/2}  \label{Roft}
\end{equation}%
Using the measured initial condition $R_{0}=3$ $\mu $m at $t=t_{0}=0$ then
yields the solid curve shown in Figure \ref{lowsigdata}. The value of $%
\sigma _{center}$ was adjusted slightly to fit the data, consistent with
measured run-to-run variations of about $\pm 15$ percent. Note that the
calibrated $\sigma _{center}$ obtained in \cite{kglcyl} includes effects
from crystal heating and incorporates (to a reasonable approximation)
differences between an infinitely long cylinder and a more realistic
half-infinite cylinder.

Observing that the solid curve provides a good fit to the measured $R(t)$
data in Figure \ref{lowsigdata} supports our initial assumption that $\alpha
_{vicinal}\gg \alpha _{diffcyl}.$ This is not a surprising result, because $%
\alpha _{diffcyl}\approx 0.003$ is quite small compared to our expectation
for a vicinal (unfaceted) surface. However, we can say little more about $%
\alpha _{vicinal}$ from the $R(t)$ data. If $\alpha _{vicinal}\gg \alpha
_{diffcyl},$ then the radial growth is strongly diffusion-limited, so $R(t)$
is essentially independent of $\alpha _{vicinal}$.

Modeling $H(t)$ cannot be done analytically, as the area of the basal
surface at the top of the column is small, and the growth is not limited
entirely by diffusion. Instead we use the cylindrically symmetric cellular
automata method described in \cite{kglca13} to numerically solve the
diffusion equation. This treats the hexagonal column as a cylindrical column
of finite length, thus allowing us to model the tip growth. 

Previous data described in \cite{kglalphas13} provide a measurement of the
basal attachment coefficient, giving $\alpha _{basal}\approx \exp (-\sigma
_{0,basal}/\sigma _{surface}),$ where $\sigma _{surface}$ is the
supersaturation at the basal surface and $\sigma _{0,basal}\approx 0.8$
percent at $T=-5$ C. We therefore tried models with this same functional
form, but with $\alpha _{0,basal}=$ 0.1, 0.2, 0.4, and 0.8 percent, yielding
the model curves for $H(t)$ shown in Figure \ref{lowsigdata}. (The model
curves for $R(t)$ are not shown in Figure \ref{lowsigdata}, as they were all
nearly identical to the analytic result for $R(t)$.) Other model parameters
included: $\alpha _{vicinal}=0.1$ and $\sigma _{out}=0.3$ percent on the
outer boundary at $R_{out}=75$ $\mu $m. (Note that $\sigma _{out}<\sigma
_{center}$ because $R_{out}<R_{far}$ and $\sigma _{center}=\sigma (R_{far})$ 
\cite{kglcyl}.)

As can be seen in Figure \ref{lowsigdata}, the $H\left( t\right) $ data are
in good agreement with the previously measured $\sigma _{0,basal}\approx 0.8$
percent. We also explored models covering a sensible range of other
parameters, again concluding that the data are consistent with $\sigma
_{0,basal}\approx 0.8$ percent, supporting the previous measurements of $%
\alpha _{basal}$ in \cite{kglalphas13}. However, the fact that $\alpha
_{vicinal}$ is not well determined limits our ability to measure $\alpha
_{0,basal}$ with high accuracy. 

The astute reader may notice that the $H(t)$ data in Figure \ref{lowsigdata}
trend upward significantly starting at $t\approx 300$ seconds, which
coincides with the formation of faceted prism surfaces. Indeed, the growth
behavior does undergo a transition with the appearance of prism facets,
indicating $\alpha _{prism}\ll \alpha _{vicinal}.$ This phenomenon is better
seen at higher growth rates, which we examine next.

\begin{figure}[htb] 
  \centering
  \includegraphics[width=3.71in,keepaspectratio]{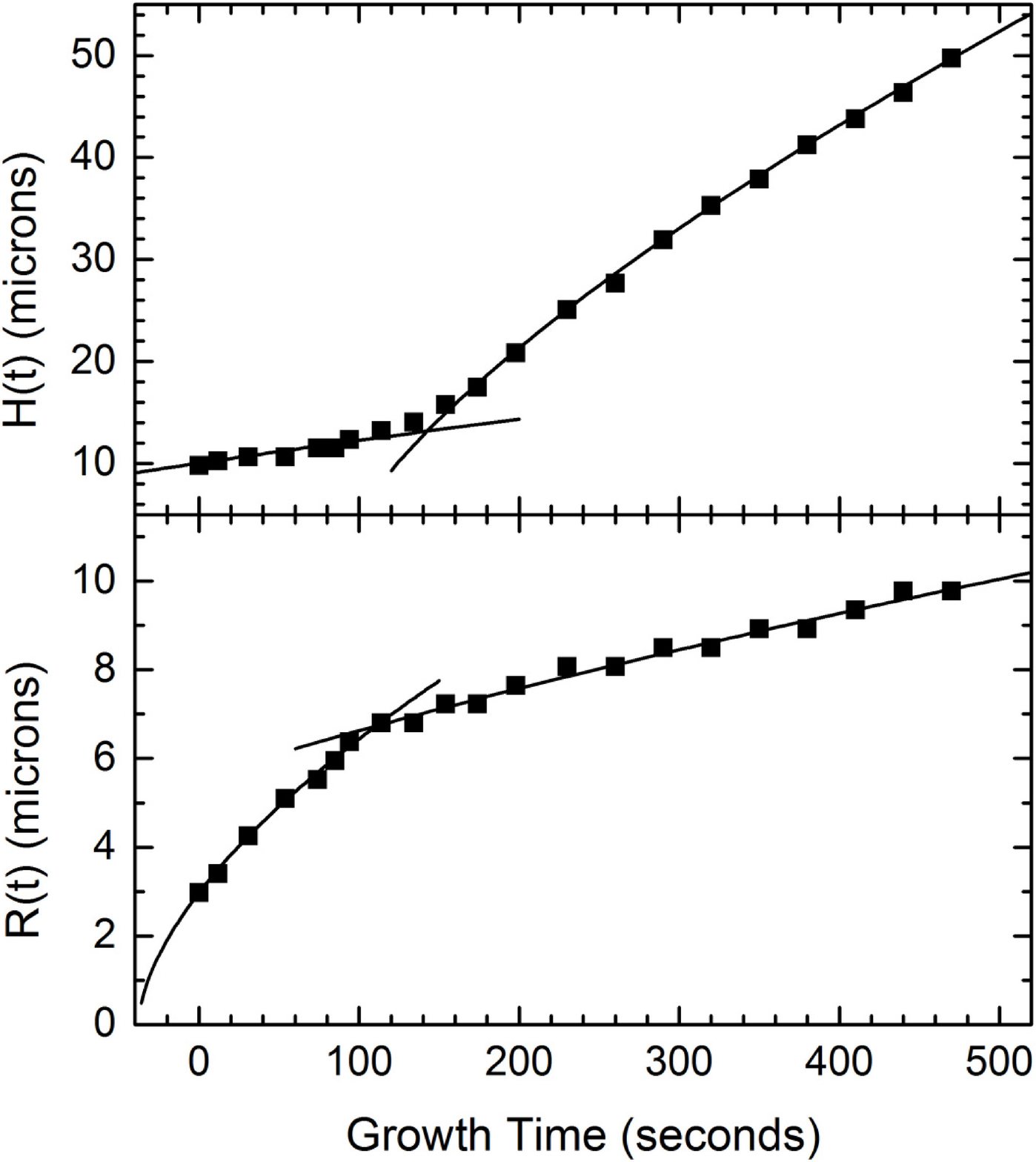}
  \caption{Measurements of the growth of
an ice needle at a supersaturation of $\protect\sigma _{center}\approx 1.8$
percent, showing the length of the needle $H(t)$ (top panel) and the needle
tip radius $R(t)$ (lower panel) as a function of growth time. A constant
length was subtracted from the $H(t)$ measurements to obtain the (arbitrary)
starting point $H(t=0)\approx 10$ $\protect\mu $m. Curves were drawn through
the data to guide the eye. Note the rather abrupt change in the growth
behavior at $t\approx t_{transition}=130$ seconds.}
  \label{set8data}
\end{figure}

\subsection{Transitional Growth at Intermediate Supersaturation}

The needle growth behavior at $-5$ C becomes more intriging when the
supersaturation is increased, and we next examine a crystal grown at $\Delta
T=3.5$ C, giving $\sigma _{center}\approx 1.8$ percent at the location of
the tip of the needle. The raw images look quite similar to the previous
data set (shown in Figure \ref{lowsigimages}) with an initially tapered
needle and no hollowing at the solid columnar tip for the duration of the
run.

Measurements obtained from the images are shown in Figure \ref{set8data},
revealing a rather abrupt transition in the growth behavior occurring in the
interval $t\approx 100-160$ seconds, during which the radial growth rate $%
dR/dt$ diminishes by about a factor of two while the axial growth rate $dH/dt
$ increases by a factor of four. Moreover, the raw images show that the
needle exhibits a clear positive taper for $t<100$ seconds, while the
columnar sides appear to be faceted prism surfaces soon thereafter.

\begin{figure}[htb] 
  \centering
  \includegraphics[width=3.58in,height=2.94in,keepaspectratio]{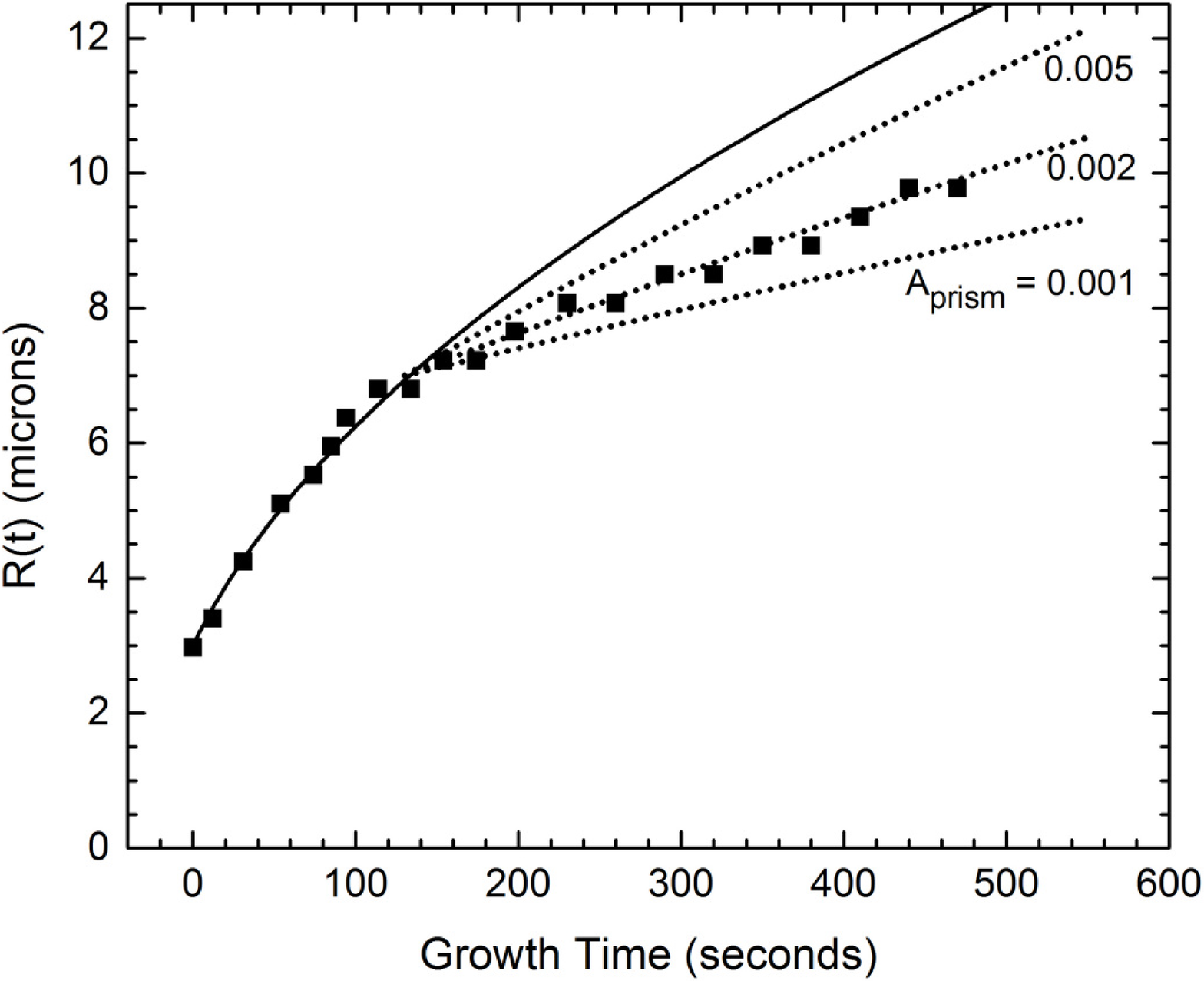}
  \caption{This graph shows the $R(t)$
data in Figure \protect\ref{set8data} along with several calculated models.
The solid curve shows Equation \protect\ref{Roft} using the measured initial
condition $R_{0}=R(t_{0})=3$ $\protect\mu $m and a small adjustment of $%
\protect\sigma _{center}$ to fit the data. This curve describes the data for 
$t<130$ seconds, when the prism surfaces are not faceted and therefore $%
\protect\alpha _{vicinal}\gg \protect\alpha _{diffcyl}.$ At later times the
prism surfaces are faceted and we cannot assume $\protect\alpha _{prism}\gg 
\protect\alpha _{diffcyl}$. The dotted curves show three numerical models
described in the text, with different $A_{prism}$ values as labeled.}
  \label{set8Rmod}
\end{figure}

For $t<t_{transition}\approx 130$ seconds, the observed needle taper
suggests that the radial growth is described by $\alpha _{vicinal}\gg \alpha
_{diffcyl}$, as we assumed in the low-$\sigma _{center}$ crystal above.
Using this assumption, we can again calculate the radial growth using
Equation \ref{Roft}, giving the solid curve shown in Figure \ref{set8Rmod}.
The good fit to the data for $t<t_{transition}$ is consistent with our
assumption that $\alpha _{vicinal}\gg \alpha _{diffcyl}\approx 0.003$, but
otherwise gives us little additional information about the magnitude of $%
\alpha _{vicinal}.$ The value of $\sigma _{center}$ needed to fit the data
was consistent with expectations from our calibration measurements \cite%
{kglcyl}. 

The fact that the $t>t_{transition}$ data deviate from the analytic $R(t)$
model (solid curve) in Figure \ref{set8Rmod} indicates that something
changed at $t\approx t_{transition}$. Because the sides of the needle became
faceted at about this time, the data indicate that the attachment
coefficient on a faceted prism surface does \textit{not} satisfy the
inequality $\alpha _{prism}\gg \alpha _{diffcyl}.$ Had this inequality been
true, the data would have followed the solid curve for $t>t_{transition}.$

When faceted prism surfaces are present for $t>t_{transition}$, we can again
make an analytic model of the radial growth rate using Equation \ref%
{diffcyl2} with $\alpha _{cyl}=\alpha _{prism},$ giving 
\begin{equation}
\frac{dR}{dt}=\frac{\alpha _{diffcyl}\alpha _{prism}}{\alpha
_{diffcyl}+\alpha _{prism}}v_{kin}\sigma _{center}  \label{diffcyl2a}
\end{equation}%
Comparing this with Equation \ref{diffcyleqn}, we see immediately that the
approximate 2x drop in $dR/dt$ at $t\approx t_{transition}$ indicates that $%
\alpha _{prism}\approx \alpha _{diffcyl}.$

Choosing the slightly model-dependent form $\alpha _{prism}(\sigma
_{surface})=A_{prism}\exp (-\sigma _{0,prism}/\sigma _{surface})$ (for
reasons we discuss below), where $A_{prism}$ and $\sigma _{0,prism}$ are
constants with $\sigma _{0,prism}=0.17$ percent \cite{kglalphas13}, we
obtained $\sigma _{surface}$ by solving the equation%
\[
\sigma _{surface}=\frac{\alpha _{diffcyl}}{\alpha _{diffcyl}+\alpha
_{prism}(\sigma _{surface})}\sigma _{center}
\]%
by iteration, which then allowed us to solve Equation \ref{diffcyl2a} for $%
R(t)$. Using different assumptions for $A_{prism}$ gave the dotted curves
shown in Figure \ref{set8Rmod}, giving a best-fit value of $A_{prism}=0.002$%
. Note that using a constant $\alpha _{prism}=0.002$ gives quite similar
results, because the final model yields that $\sigma _{surface}$ is roughly
six times larger than $\sigma _{0,prism},$ and therefore $\exp (-\sigma
_{0,prism}/\sigma _{surface})\approx 1.$

\begin{figure}[htb] 
  \centering
  \includegraphics[width=3.75in,height=3.07in,keepaspectratio]{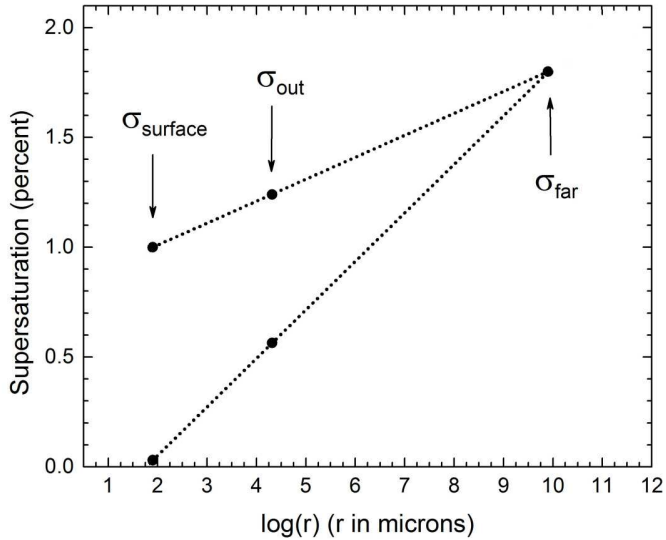}
  \caption{This plot shows the model
supersaturation $\protect\sigma (r)$ surrounding the columnar crystal as a
function of $\log (r)$, as described in the text.  For $t<t_{transition}$, $%
\protect\alpha _{vicinal}\gg \protect\alpha _{diffcyl}$ and we obtain the
lower dotted line, using the known outer boundary condition $\protect\sigma %
(R_{far})=\protect\sigma _{far}\approx 1.8$ percent at $R_{far}\approx 2$
cm. For $t>t_{transition}$, the columnar walls are faceted with a much lower 
$\protect\alpha _{prism}\approx 0.002,$ yielding the upper line shown here.}
  \label{sigmamod}
\end{figure}

It is instructive at this point to examine the supersaturation field $\sigma
(r)$ around the growing crystal, again using the infinite-cylinder
approximation. The general solution to the diffusion equation around an
infinite cylinder is $\sigma \left( r\right) =A_{1}+A_{2}\log (r),$ where $%
A_{1}$ and $A_{2}$ are constants determined by the boundary conditions. The
two models above for $t<t_{transition}$ and $t>t_{transition}$ give the two $%
\sigma \left( r\right) $ lines shown in Figure \ref{sigmamod}, where we have
indicated the special radii $r=R_{in}=7$ $\mu $m (the crystal surface at $%
t=t_{transition}),$ $r=R_{out}=75$ $\mu $m (the outer boundary used in our
numerical models), and $r=R_{far}=2$ cm (the effective outer boundary in our
analytic models, as determined by the supersaturation calibration in \cite%
{kglcyl}).

The supersaturation at $R_{far}$ is given by $\sigma _{far}=\sigma
_{center}\approx 1.8$ percent, which is fixed by the temperature profile of
the diffusion chamber. Because this temperature profile remained constant as
the crystal grew, the value of $\sigma _{far}$ does not change at $%
t=t_{transition}$ when the sides of the needle became prism facets.

In contrast, the value of $\sigma _{surface}$ depends on the inner boundary
condition at $r=R_{in}$, which does change at $t=t_{transition}.$ For times $%
t<t_{transition},$ we have $\sigma _{surface}\approx \left( \alpha
_{diffcyl}/\alpha _{vicinal}\right) \sigma _{center},$ coming from $%
dR/dt=\alpha _{diffcyl}v_{kin}\sigma _{center}=\alpha
_{vicinal}v_{kin}\sigma _{surface},$ which is one form of the inner boundary
condition. The fact that $\alpha _{vicinal}\gg \alpha _{diffcyl}$ means that 
$\sigma (r=R_{in})$ is quite small for $t<t_{transition}$, as shown in
Figure \ref{sigmamod}. Note that the straight lines in the figure connect
the known $\sigma $ values at $r=R_{in}$ and $r=R_{far}$ to give the full
solutions $\sigma \left( r\right) =A_{1}+A_{2}\log (r).$

For $t>t_{transition},$ our best fit $A_{prism}=0.002$ combines with $\alpha
_{diffcyl}$ to yield $\sigma _{surface}\approx 1$ percent, yielding the
upper dotted line in Figure \ref{sigmamod}. A key point in Figure \ref%
{sigmamod} is that the drop in $\alpha $ on the sides of the column from $%
\alpha _{vicinal}$ to $\alpha _{prism}$ at $t\approx t_{transition}$ results
in a remarkably large jump in $\sigma _{surface},$ from near zero to about
one percent. It is this sudden change in $\sigma _{surface}$ that causes the
observed jump in $dH/dt$.

\begin{figure}[htb] 
  \centering
  \includegraphics[width=3.5in,keepaspectratio]{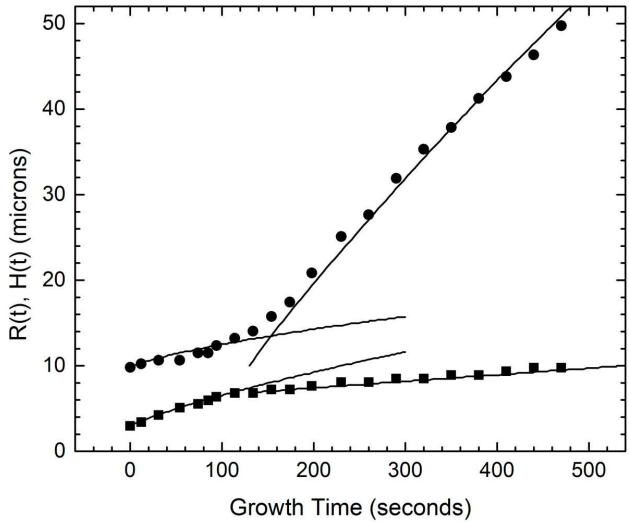}
  \caption{This graph shows the data in
Figure \protect\ref{set8data} along with two numerical models, one for $%
t<t_{transition}=130$ seconds (modeling both $R(t)$ and $H(t)$), and a
second model for $t>t_{transition}.$ Using the basal attachment coefficient
measured in \protect\cite{kglalphas13}, together with the parameters
determined by our analytic modeling of $R(t),$ yields good agreement with
both the $R(t)$ and $H(t)$ data. The change at $t\approx t_{transition}$
arises when the sides of the columnar crystal become faceted, changing the
attachment coefficient from $\protect\alpha _{vicinal}$ to $\protect\alpha %
_{prism}\approx 0.002.$ This slows the radial growth rate $dR/dt,$ but also
causes a large jump in $\protect\sigma _{surface},$ as shown in Figure 
\protect\ref{sigmamod}. This jump then causes the axial growth rate $dH/dt$
to increase by about a factor of four.}
  \label{camodels}
\end{figure}

To model $H(t),$ and to deal with the free end of the column more generally
(no longer in the infinite-cylinder approximation) we again use the
cylindrically symmetrical cellular automata model described in \cite{kglca13}%
, and the model results are shown with the data in Figure \ref{camodels}.
Importantly, these models were not adjusted to fit the data, as all the
model inputs were determined by other measurements. For $t<t_{transition},$
we used $\alpha _{vicinal}\approx 0.1$ (the exact value was not important as
long as $\alpha _{vicinal}\gg \alpha _{diffcyl}$) and $\alpha _{basal}=\exp
(-\sigma _{0,basal}/\sigma _{surf})$ with $\sigma _{0,basal}=0.8$ percent,
as measured in \cite{kglalphas13}. For $t>t_{transition}$ we used $%
A_{prism}=0.002$ from our analytical modeling. The supersaturation at the
outer boundary at $r=R_{out}=75$ $\mu $m changed from $\sigma _{out}=0.6$
percent for $t\,<t_{transition}$ to $\sigma _{out}=1.0$ percent for $%
t>t_{transition}$ seconds, with these values provided by the analytic models
shown in Figure \ref{sigmamod}.

\begin{figure}[htb] 
  \centering
  \includegraphics[width=3.5in,keepaspectratio]{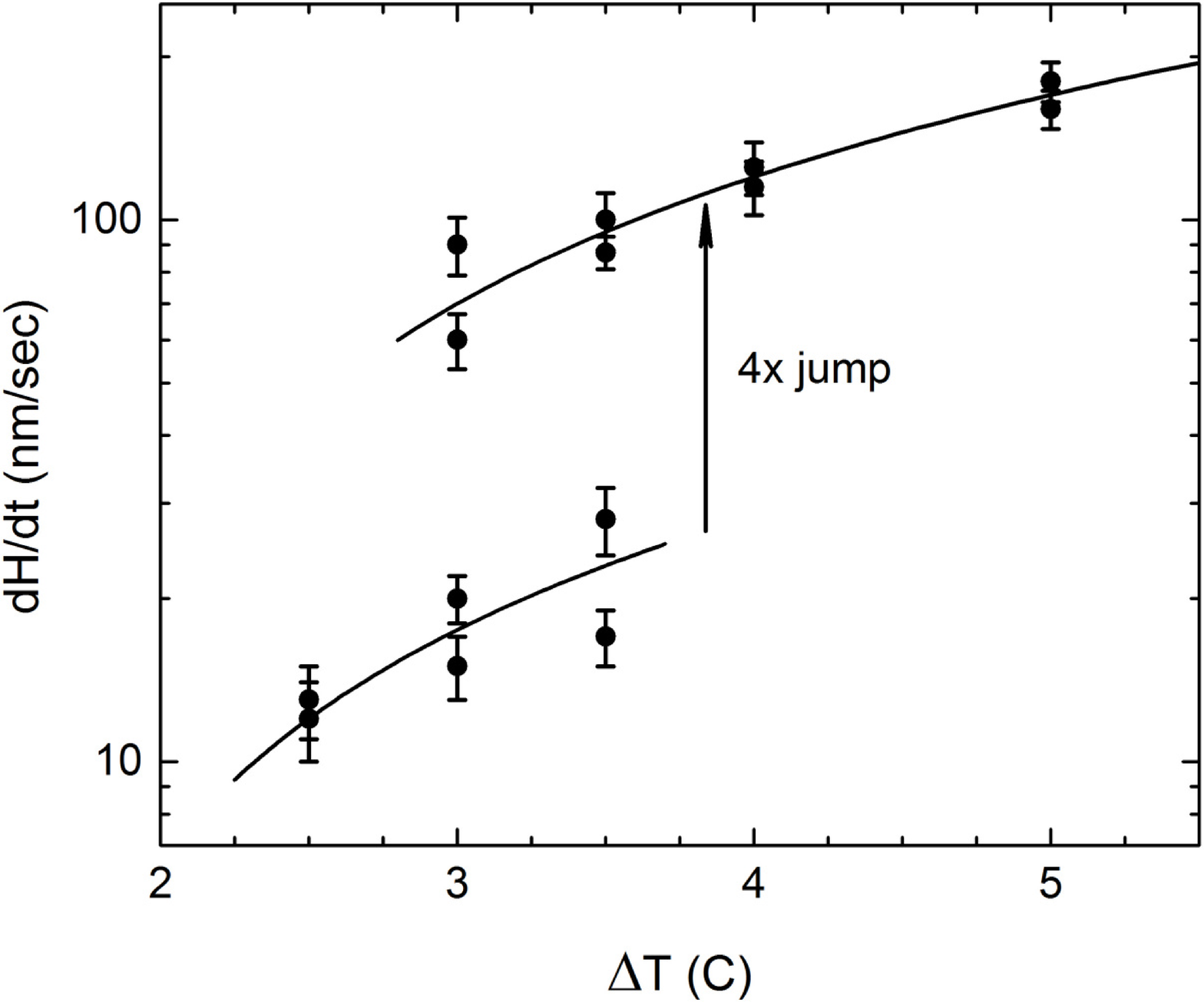}
  \caption{This graph shows the axial
growth rate $dH/dt$ for several needle crystals covering a range of $\Delta
T.$ For $\Delta T=3$ and 3.5 C, $dH/dt$ was measured both before and after
the transition caused by faceting of the prism surfaces. At higher and lower 
$\Delta T,$ no jump in $dH/dt$ was observed.}
  \label{hjump}
\end{figure}

Our essential conclusion from the $H(t)$ modeling shown in Figure \ref%
{camodels} is a simple one: the basal attachment coefficient $\alpha
_{basal}=\exp (-\sigma _{0,basal}/\sigma _{surf})$ from the low-pressure
data in \cite{kglalphas13} provides good agreement with the needle growth
data taken in air. Figure \ref{hjump} shows that the transitional jump in $%
dH/dt$ was observed only at intermediate supersaturations.

\subsection{Thermal Effects}

It is also worth pointing out that thermal effects from latent heating are
beginning to become significant at this intermediate supersaturation level.
In the infinite-cylinder approximation, latent heating yields a temperature
increase of the ice (relative to the far-away air temperature) of \cite%
{kglcyl} 
\[
\delta T=\frac{B\lambda \rho vR}{\kappa }
\]%
which gives $\delta T\approx 0.2$ C for $t\,<t_{transition}$ and $\delta
T\approx 0.06$ C for $t\,>t_{transition}$, not including additional heating
from the axial growth. Heating from $dH/dt$ is more difficult to estimate,
as it is concentrated at the needle tip, so heat is conducted down the
needle and then into the surrounding air. Nevertheless, we estimate that
heating effects in this crystal are dominated by heating from radial growth,
as the tip area is quite small. While we did not solve the full
particle+heat diffusion problem, our approximate analysis suggests that
heating created a relatively small correction to the above analysis, and
does not alter our conclusions.

\subsection{Growth at High Supersaturation}

Raising $\sigma _{center}$ still higher, Figure \ref{example} shows an
additional test crystal grown at $\Delta T=5$ C, for which $\sigma
_{center}\approx 3.7$ percent. At this higher supersaturation, the faceted
columnar morphology includes a conical hollowed structure at the tip, which
developed as the needle grew longer. Moreover, the initial positive taper of
the column almost immediately reversed to an overall negative taper, with $R$
largest at the tip, as is seen in Figure \ref{example}. Measurements of $R(t)
$ show a simple $R\sim t^{1/2}$ behavior, with no obvious transitions. The
observations thus indicate that $\alpha _{prism}>\alpha _{diffcyl}\approx
0.002$ because the top prism terrace at the end of the needle is growing at
essentially the same velocity as the vicinal surfaces making up the negative
taper. Beyond this inequality, however, we cannot determine $\alpha _{prism}$
from the columnar growth.

At still higher $\sigma _{center}$, the prism facets disappear entirely, and
the morphology transitions into \textquotedblleft
fishbone\textquotedblright\ dendrites \cite{fishbones09, kgldual14} for $%
\sigma _{center}>15$ percent. The disappearance of prism facets indicates $%
\alpha _{prism}\approx 1$ at these high supersaturations. However, heating
plays a larger role at higher supersaturations as well, making it necessary
to better incorporate heating effects into our growth models before we can
draw reliable conclusions.

\section{Discussion}

Our principal conclusion from this work is that $\alpha _{prism}\approx 0.002
$ at temperatures near -5 C in air at $p_{air}=1$ bar. This statement
strictly applies for a surface supersaturation near $\sigma
_{surface}\approx 1$ percent, with an overall uncertainty in $\alpha _{prism}
$ of roughly a factor of two. Similar values of $\alpha _{prism}$ were
obtained over a larger range of $\sigma _{surface}$ in \cite%
{libbrechtarnold09}, in general agreement with the current work. We believe
these data provide the most accurate quantitative determinations of $\alpha
_{prism}$ to date, supporting many decades of morphological observations of
the growth of slender columnar ice crystals in air near -5 C.

This exceptionally low value of $\alpha _{prism}$ is strongly inconsistent
with our previous measurment of $\alpha _{prism}\approx 0.15\exp (-\sigma
_{0,prism}/\sigma _{surface})$ with $\sigma _{0,prism}\approx 0.17$ percent,
which we obtained at $p_{air}\approx 0.01$ bar. We believe that both these
measurements are accurate, as both were done in well controlled conditions,
giving considerable attention to diffusion modeling and eliminating
systematic errors. The clear descrepancy in the measurements thus forces us
to the conclusion that $\alpha _{prism}$ depends on $p_{air}$, dropping by
nearly two orders of magnitude as $p_{air}$ is increased from 0.01 to 1 bar.

There have been other indications that $\alpha _{prism}$ is quite small near
-5C, and others have speculated that the ice attachment coefficients might
depend on $p_{air}$. Howver, interpreting many of the older ice-growth
observations into a quantitative measure of $\alpha _{prism}$ has been
problematic. Systematic errors, especially relating to precise modeling of
particle diffusion, have made it difficult to accurately relate growth rates
to attachment coefficients \cite{critical04, substrateinteractions12}. In
contrast, the growth transition of the intermediate-$\sigma _{center}$
crystal presented above makes an especially strong case for $\alpha
_{prism}\approx 0.002.$

In a previous report we incorrectly assumed that $\alpha _{prism}$ was not
affected by $p_{air}$, which then led us to the conclusion that changes in $%
\alpha _{basal}$ were necessary to explain the growth of columnar crystals
at -5 C \cite{SDAK5}. As we stated in \cite{SDAK5}, this assumption was
consistent with the data available at that time, but we also wrote that we
could not \textquotedblleft positively exclude that there may be some
pressure dependence in $\alpha _{prism}$.\textquotedblright\ The new results
presented here negate this important assumption made in \cite{SDAK5}, and
thus negate our conclusion that changes in $\alpha _{basal}$ with terrace
thickness are required to explain the growth of columnar crystals near -5 C.
The data now suggest that no changes in $\alpha _{basal}$ with terrace
thickness are necessary, and further suggest that $\alpha _{basal}$ does not
change with $p_{air}$. However, the data still support our hypothesis that $%
\alpha _{prism}$ does change with terrace thickness near -15 C \cite{sdak12,
esi15}.

We can offer no microscopic physical model to explain the measured $\alpha
_{prism}(\sigma _{surface},p_{air})$, especially the dependence on $p_{air}$%
. We considered the possibility that $\sigma _{0,prism}$ increased with $%
p_{air},$ as this could yield the small $\alpha _{prism}$ values measured.
Such a model would yield strong changes in $\alpha _{prism}$ as a function
of $\sigma _{surface},$ however, and this behavior seems to be excluded by
other measurements \cite{libbrechtarnold09}. Moreover, we are inclined to
think, on physical grounds, that the presence of air at the crystal surface
should not change the step energy of a prism terrace, and thus should not
change $\sigma _{0,prism}$ from the value $\sigma _{0,prism}\approx 0.17$
percent measured at $p_{air}\approx 0.01$ bar.

Given these considerations, we suggested the separated functional form%
\[
\alpha _{prism}=A_{prism}(\sigma _{surface},p_{air})\exp (-\sigma
_{0,prism}/\sigma _{surface})
\]
described above, keeping $\sigma _{0,prism}\approx 0.17$ percent. The data
then suggest $A_{prism}\approx 0.002$ for $\sigma _{surface}\approx 1$
percent, and that $A_{prism}$ increases with higher $\sigma _{surface}$,
eventually becoming $A_{prism}\approx 1$ when $\sigma _{surface}\gg 1$
percent. This low-$\sigma _{0,prism}$ model makes a prediction that, for very
low $\sigma _{surface}$, we should find 
$\alpha_{prism}\approx 0.002\exp (-\sigma _{0,prism}/\sigma _{surface})$,
and this could be tested with additional measurements.
Without a better understanding of the molecular processes underlying $\alpha
_{prism},$ or more definative measurements, we cannot carry the discussion
much further.

These new revelations change our overall picture of ice growth dynamics. The
data now suggest that $\alpha _{basal}$ is described simply by the
measurements in \cite{kglalphas13}, with no depedence on $p_{air},$ and no
dependence on terrace width. All of the peculiar behavior is now placed on
the prism facet, with $\alpha _{prism}$ depending strongly on $p_{air}$ at $%
T=-5$ C, with a complex dependence on $\sigma _{surface}$ as well. Moreover,
we have found in other measurments that $\alpha _{prism}$ depends strongly
on terrace width near $T=-15$ C. All this peculiar behavior may be related
to the onset of surface roughening on the prism facet, which appears to
happen gradually over the temperature range $-1>T>-10$ C \cite{kglalphas13}.

Although we are making progress toward a comprehensive model of ice growth,
our picture remains phenomenologically complex. We still possess little real
understanding of the fundamental molecular processes responsible for the
observed behavior of the attachment coefficients with temperature,
supersaturaion, and background air pressure.

\bibliography{C:/Dropbox/1-kgl-top/Papers/1Bibliography/kglbiblio3}
\bibliographystyle{phreport}

\end{document}